\documentclass[paper]{ieice}
\usepackage{graphicx,xcolor}
\usepackage{amssymb,amsmath,amsthm,amsfonts,bm,bbm,mathtools,booktabs,cite}
\usepackage{algorithm,algpseudocode}
\usepackage{optidef}
\usepackage{subfigure}
\usepackage[cmintegrals]{newtxmath}
\usepackage{tabularx}
\usepackage{multirow}
\usepackage{url}

\algnewcommand\Initialize{\textbf{Initialize: }}

\DeclareMathOperator*{\argmax}{arg\,max}
\def\ind{\mathop{\mathbbm{1}}\nolimits}
\theoremstyle{definition}
\newtheorem{remark}{Remark}

\graphicspath{{./_figs/}}

\field{B}
\title{Penalized and Decentralized Contextual Bandit Learning for WLAN Channel Allocation with Contention-Driven Feature Extraction}
\authorlist{%
\authorentry{Kota Yamashita}{s}{labelA}\MembershipNumber{2009122}
\authorentry{Shotaro Kamiya}{n}{labelB}\MembershipNumber{}
\authorentry{Koji YAMAMOTO}{e}{labelA}\MembershipNumber{0212512}
\authorentry{Yusuke Koda}{n}{labelC}\MembershipNumber{}
\authorentry{Takayuki NISHIO}{m}{labelD}\MembershipNumber{1002899}
\authorentry{Masahiro Morikura}{f}{labelA}\MembershipNumber{}
}
\affiliate[labelA]{The authors are with the Graduate School of Informatics, Kyoto University Yoshida-honmachi, Sakyo-ku, Kyoto 606-8501, Japan.}
\affiliate[labelB]{The author is with Sony Corporation, 1-7-1 Konan Minato-ku, Tokyo, 108-0075, Japan.}
\affiliate[labelC]{The author is with the Centre for Wireless Communications, University of Oulu, 90014, Finland,}
\affiliate[labelD]{The author is with School of Engineering, Tokyo Institute of Technology, Ookayama, Meguro-ku, Tokyo, 158-0084, Japan.}

\received{2021}{11}{19}
\revised{2021}{1}{19}

\begin{document}
\maketitle
\begin{summary}
    In this study, a contextual multi-armed bandit (CMAB)-based decentralized channel exploration framework disentangling a channel utility function (i.e., reward) with respect to contending neighboring access points (APs) is proposed.
    The proposed framework enables APs to evaluate observed rewards compositionally for contending APs, allowing both robustness against reward fluctuation due to neighboring APs' varying channels and assessment of even unexplored channels.
    To realize this framework, we propose contention-driven feature extraction (CDFE), which extracts the adjacency relation among APs under contention and forms the basis for expressing reward functions in the disentangled form, that is, a linear combination of parameters associated with neighboring APs under contention).
    This allows the CMAB to be leveraged with joint a linear upper confidence bound (JLinUCB) exploration and to delve into the effectiveness of the proposed framework. 
    Moreover, we address the problem of non-convergence---the channel exploration cycle---by proposing a penalized JLinUCB (P-JLinUCB) based on the key idea of introducing a discount parameter to the reward for exploiting a different channel before and after the learning round.
    Numerical evaluations confirm that the proposed method allows APs to assess the channel quality robustly against reward fluctuations by CDFE and achieves better convergence properties by P-JLinUCB.
\end{summary}

\section{Introduction}
\label{sec:introduction}
Owing to the rapid development of the Internet of things (IoT) technology, 
the number of access points (APs) in wireless local area networks (WLANs) is steadily increasing\cite{cisco}.
In environments where APs are densely deployed, the transmission opportunity of each AP is limited.
This is because the IEEE 802.11 standard for WLANs is based on the carrier sense multiple access with collision avoidance (CSMA/CA) protocol as a medium access control (MAC) technique.
Furthermore, with an increase in applications such as online video conferencing and cloud computing, the requirements for high throughput and low latency have become more demanding\cite{wifi7}.
To meet these requirements, a next-generation WLAN technology, called IEEE 802.11be, is being discussed.
The objectives of IEEE 802.11be are to \textit{1)} enable a new MAC and physical (PHY) mode operation that can support a maximum throughput of at least 30\,Gbit/s 
and \textit{2)} ensure backward compatibility and coexistence with legacy 802.11 devices in the unlicensed bands of 2.4, 5, and 6\,GHz \cite{11be}.
Despite the densely deployed WLAN environment, the number of available channels is limited, which requires a new channel allocation method to achieve high throughput.

Our interest is in decentralized channel allocation in unknown WLAN environments (i.e., when the expected throughput of each channel is unknown).
However, in general, for APs using the same channel that perform time-division transmission, 
the resultant throughput is not necessarily deterministic because the conditions of the neighboring APs (e.g., traffic) vary at each instant and cannot be known in advance.
This observation suggests that it is necessary to devote efforts to information collection online.
Therefore, we need to successively explore a channel while aggregating information and finally exploit the optimal channel.

The above strategy can be formulated as a multi-armed bandit (MAB) problem\cite{ucb, mab,bandit_algorithm}.
In the context of WLANs, only a few recent studies\cite{compare_wlan_2019,compare_wlan_2020} have applied the MAB technique to the channel allocation problem.
These studies demonstrate that the action selection strategy of an agent (e.g., AP or station (STA)) according to the MAB algorithm is beneficial for resource allocation problems, including channel allocation in WLANs.
It should be noted that generally the quality in a channel-wise metric, that is, the channel reward, significantly changes when another AP changes its channel. 
Nonetheless, the above mentioned studies only estimate the expected channel reward being agnostic to the selected channels of the neighboring APs.
Hence, in the aforementioned studies, the agent needs to re-experience the channel to track the updated reward expectation of the channel, which reduces the exploration efficiency.

To overcome this disadvantage, we focus on the neighboring APs rather than on the channel conditions themselves. 
Our key insight is that, in general, the transmission probability of an AP is constant regardless of the channel selected by the AP. 
Therefore, if a neighboring AP has selected the same channel as an AP of interest at least once, 
the latter AP should be able to leverage this knowledge to predict the fluctuation in the expected channel reward owing to the channel adjustment of the neighboring AP. 
This prediction disentangled with respect to neighboring APs enables accurate estimation of the expected reward of a channel with a small number of channel explorations or without any channel observation. 
This enables us to make a better decision (i.e., whether to change the channel or not) and thus improve the throughput of the system.


\begin{table*}[t]
  \caption{Comparison between this paper and related works on decentralized channel allocation based on MAB learning for wireless networks.}
  \label{tab:related_works}
  \centering
 \begin{tabularx}{\textwidth}{cccccc}
  \toprule
  \textbf{Reference} &\textbf{Method} &\textbf{Interference} &\textbf{Observability}&\textbf{WLAN?} \\
  \midrule
  \cite{compare_2014_1}&Adversarial MAB&Full interference&Local reward&No\\
  \cite{compare_2014_2}&MAB \& Calibrated forecaster\cite{calibrated_forecaster}&Full interference&Local reward \& neighbor's action&No\\ 
  \cite{compare_2016_1}&MAB&Graph-based&Local reward \& neighbor's information\footnotemark&No\\
  \cite{compare_wlan_2019}&MAB&Graph-based&Local reward&Yes\\
  \cite{compare_wlan_2020}&MAB&Graph-based&Local reward&Yes\\
  This paper&CMAB&Graph-based&Local reward \& neighbor's action&Yes\\
  \bottomrule
 \end{tabularx}
\end{table*}

\begin{figure}[t]
  \centering
  \includegraphics[width=\columnwidth]{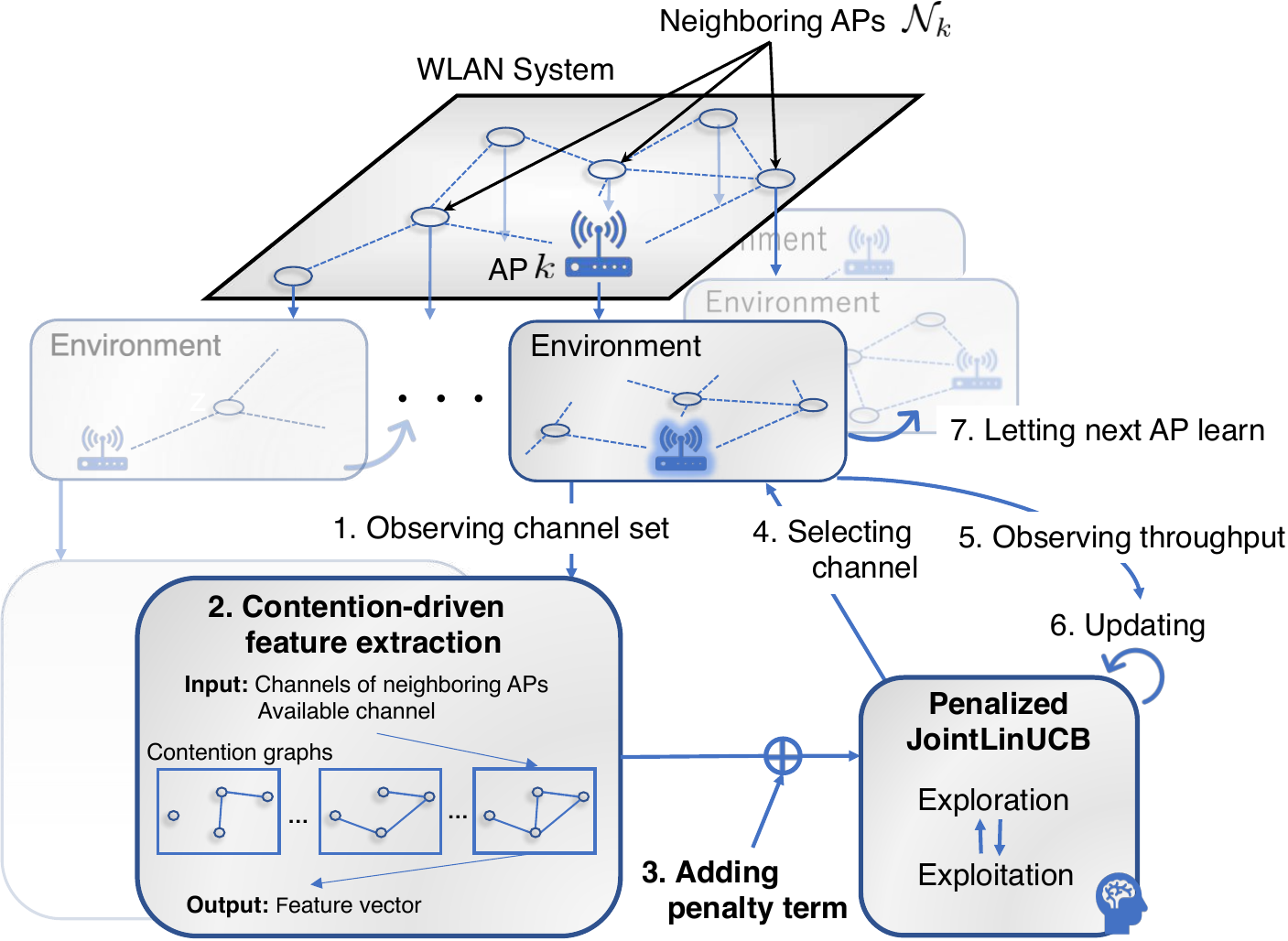}
  \caption{Overview of the proposed decentralized WLAN channel allocation method incorporating prior information of neighboring APs.}
  \label{fig:proposed_scheme}
\end{figure}

Based on the idea mentioned above, we formulated a novel method of WLAN channel selection that considers the past transmission probabilities of other APs in the same channel, using a contextual MAB (CMAB) framework\cite{linucb1,linucb2}. 
Fig.~\ref{fig:proposed_scheme} outlines the proposed scheme.
Specifically, we incorporate the channels selected by neighboring APs as prior information (i.e., context) into the linear CMAB learning. 
As this framework typically requires the design of problem-specific feature vectors using the context (see Section \ref{sec:preliminaries}), we construct a contention-driven feature extraction (CDFE) scheme for WLAN channel selection based on the CMAB algorithm.
Focusing on the fact that the communication quality in WLANs depends heavily on the contention graph\cite{csrange1}, 
CDFE uses the channels of the neighboring APs as the input and outputs the feature vectors corresponding to the adjacency relation of the contention graph. 
Unlike simple MAB learning, CMAB learning with CDFE selects the channel to reach the contention graph with the highest expected channel reward.
This enables accurate channel reward estimation even for unexplored channels as long as neighboring APs and their conditions remain constant, thus allowing for a more efficient channel exploration.


Meanwhile, selfish decentralized learning has an inherent problem in that the entire system does not always converge to a fixed strategy. 
This suggests the possibility of behavioral cycles in the proposed CMAB-based channel allocation. 
To address this problem, we also propose a penalized JointLinUCB (P-JLinUCB), which is an extension of LinUCB\cite{linucb1,linucb2}. 
The proposed P-JLinUCB introduces a parameter that discounts the reward observed when the channel is varied and adds a term corresponding to the penalty to the feature vector.
This added term realizes the penalty for a particular action in linear CMAB learning.
Consequently, P-JLinUCB reduces the variability in channel allocation.

\subsection{Related Works}
\label{subsec:relatedworks}
The MAB algorithm is categorized as reinforcement learning (RL), which focuses on balancing the exploration–exploitation trade-offs\cite{sutton2018reinforcement} inherent in RL. 
When considering resource allocation on wireless networks, we frequently face a situation wherein we require RL to learn effective resource allocation while exploring the performance (e.g., system throughput, frame loss rate, and delay) resulting from a typical resource allocation. 
This is because the actual performance is not known in advance.
Therefore, MAB-based resource allocation has been discussed in several studies.
Modi \textit{et al.}\@ \cite{modi2017qos} proposed online learning algorithms based on MAB theory for opportunistic spectrum access by secondary users (SUs) in cognitive radio networks, 
where there is no information exchange among the SUs.
Zhou \textit{et al.}\@ \cite{zhou2020human} focused on human behavioral data (e.g., user location, quality of experience (QoE)-aware data) generated in 5G networks 
and proposed a method to exploit such data for dynamic channel allocation using a CMAB algorithm.
The MAB-based formulation is also found in other resource allocation problems \cite{gai2014distributed,deng2020thompson}.
As is evident from the above, the MAB application area covers a wide variety of resource allocation problems, and channel allocation is no exception.

Because the channel allocation problem is compatible with the MAB problem as described above, 
several MAB approaches to decentralized channel allocation have been proposed not only for WLANs, but also for other wireless networks.
Table \ref{tab:related_works} lists a brief comparison of the related studies and our study.
In \cite{compare_2014_1}, channel selection and power control in infrastructural networks were modeled as a multi-agent adversarial MAB game.
In \cite{compare_2014_2}, the channel selection problem in underlay distributed device-to-device (D2D) communication systems was considered. 
As a multiplayer MAB game, each D2D user selects a channel based on a calibrated forecaster\cite{calibrated_forecaster} and no-regret learning.
These two studies discuss the convergence of the channel allocation in detail. 
However, both these studies assume a \textit{full interference} model\cite{interference_model} in which any two or more agents interact with each other. 
This assumption does not consider the situation in which contention in WLANs is represented as a \textit{graph-based} model\cite{csrange1,interference_model}.
The authors in \cite{compare_2016_1} proposed a decentralized channel allocation method for cognitive radio networks based on a graph coloring algorithm\cite{graph_color} 
and the MAB algorithm using a graph representation of the network of SUs.
Notably, this method requires information sharing among agents.
MAB-based approaches can also be found in the context of WLAN channel allocation.
In \cite{compare_wlan_2019}, the effectiveness of well-known MAB algorithms such as the UCB algorithm\cite{ucb} and Thompson sampling\cite{thompson_sampling}, 
for channel allocation in densely deployed WLANs was examined from multiple perspectives.
In \cite{compare_wlan_2020}, the feasibility of channel assignment and AP selection using Thompson sampling, under which both APs and STAs were empowered with agents, was studied.

\footnotetext[1]{This implies that some learned information needs to be shared among agents.}
These WLAN-specific schemes do not use prior information regarding the channels selected by neighboring APs in MAB learning; thus, the benefit of the prior information for MAB-based channel allocation has not been studied.
Moreover, these schemes require a thorough exploration of the channel to estimate the expected channel reward. 
In contrast to these studies, by using prior information appropriately, this study estimates the channel reward disentangled with respect to neighboring APs from the exploration experience.
This enables the assessment of even unexplored channels because from the disentangled channel reward and prior information, the expected channel reward can be calculated via a linear combination.
It should be noted that \cite{compare_wlan_2019} assumed that the traffic is identical for all APs. 
It is also worth noting that in \cite{compare_wlan_2020}, not only APs but also STAs are learned as agents to distribute the traffic, thereby not addressing the traffic condition-wise channel allocation. 

\subsection{Our Contributions and Paper Organization}
\label{subsec:contributions}
The main contributions of this study are summarized as follows:
\begin{itemize}
  \item We propose a CMAB-based decentralized WLAN channel exploration framework that estimates \textit{disentangled channel rewards with respect to each neighboring AP}, which is robust against channel adjustments of the neighboring APs and the presence of unexplored channels. 
  This is achieved by leveraging the neighboring APs' occupied channels as prior information (i.e., context).
  The proposed framework is in stark contrast to the existing studies on MAB-based decentralized WLAN channel allocation schemes\cite{compare_wlan_2019,compare_wlan_2020}, which leveraged only channel rewards without context and lacked the perspective of the aforementioned AP-wise disentanglement of rewards.
  
  \item To realize the above disentangled channel reward estimation, we propose a feature extraction scheme for linear CMAB learning, named CDFE. 
  CDFE allows the linear CMAB algorithm to learn individual parameters representing each AP's impact on the channel reward (e.g., traffic condition), thereby forming channel allocation strategies in view of such conditions. 
  The key technique of CDFE is to leverage the selected channels of neighboring APs as contexts and design feature vectors corresponding to the adjacencies of the contention graph.

  \item One concern is that decentralized and selfish learning does not necessarily converge, i.e., cycles of channel allocations continue.
  To this end, we propose P-JLinUCB, which applies a discount parameter to rewards for specific actions.
  In P-JLinUCB, to reflect the impact of the discounted reward on the learning model, we add a penalty term to the feature vector. 
  These schemes allow the convergence probability of channel allocation to be increased without significantly degrading the performance.
  We highlight that this framework can also be applied to the case of incorporating penalties into general linear CMAB problems.
\end{itemize}

The remainder of this paper is organized as follows.
In Section~\ref{sec:preliminaries}, we describe the CMAB problem and LinUCB as a supplement to this study.
In Section~\ref{sec:system_model}, we present our system model and problem formulation.
In Section~\ref{sec:propose}, we present a decentralized channel allocation method based on LinUCB with penalties and a feature extraction method for that purpose.
In Section~\ref{sec:evaluate}, we perform numerical evaluations of the proposed channel allocation method.
In Section~\ref{sec:conclusion}, we conclude this study.

\textit{Notation: }$\mathbb{E}[\cdot]$ denotes the expectation operator, 
and $\ind(y)$ denotes the indicator function that equal to 1 if event $y$ is true and 0 otherwise.
We denote the inner product by $\langle\cdot,\cdot\rangle$. 
We let superscript $(t)$ denote the time step.
For any sequence $\{w^{(t)}\}_{t=0}^{\infty}$, 
we use $w^{(t_1:t_2)}$ to denote the sub-sequences $w^{(t_1)}, w^{(t_1+1)},\dots, w^{(t_2)}$.

\section{Preliminaries}
\label{sec:preliminaries}
This section introduces the CMAB and LinUCB principles.
\subsection{Linear Contextual Multi-Armed Bandit Problem}
\label{subsec:cmab}
This section describes the linear CMAB problem formally\cite{linear_bandit_1,linear_bandit_2,linear_bandit_3,bandit_algorithm}.
Let $\mathcal{A}$ be a finite set of arms (i.e., action set), and $\bm{x}\in\mathcal{X}$ be a context vector, where $\mathcal{X}$ is an arbitrary fixed set of context vectors, 
and $r(a) \in [0,1]$ is the reward of arm $a \in \mathcal{A}$.
In the linear CMAB setting, 
the expected reward of an arm $a$ is linear in its $d$-dimensional feature vector $\bm{\varphi}(\bm{x}, a)$ 
with some unknown coefficient vector $\bm{\theta}^{\star}\in\mathbb{R}^{d}$;
that is, it is assumed that
\begin{align}
  \label{eq:cmab_reward}
  \mathbb{E}[r(a)\,\vert\,\bm{x},a] = \langle\bm{\varphi}(\bm{x}, a), \bm{\theta}^{\star}\rangle.
\end{align}
\begin{remark}
  A map from a context-arm pair to a feature vector $\bm{\varphi}:\mathcal{X}\times\mathcal{A}\rightarrow\mathbb{R}^d$ is an arbitrary but known function\cite{linear_bandit_3} (i.e., $\bm{\varphi}$) can be freely defined by users.
  Therefore, the key to performing learning rapidly and efficiently is to construct a map $\bm{\varphi}$ suitable for the problem setting.
\end{remark}

The following steps are performed in each trial $t=1, 2,\dots, T$:
\begin{enumerate}
  \item The context vector $\bm{x}^{(t)}$ is revealed to the agent.
  \item The agent chooses an arm $a^{(t)} \in {\mathcal{A}}$ in accordance with a CMAB algorithm.
  \item The agent observes the reward $r^{(t)}(a^{(t)}) \in [0,1]$.
\end{enumerate}
In the linear CMAB setting, the observed reward $r^{(t)}(a^{(t)})$ is assumed to satisfy
\begin{align}
  \label{eq:linear_reward}
  r^{(t)}(a^{(t)}) = \langle\bm{\varphi}(\bm{x}^{(t)}, a^{(t)}), \bm{\theta}^{\star}\rangle + \eta^{(t)},
\end{align}
where $\eta^{(t)}$ is random noise, such that with a fixed constant $R\geq 0$, for any $\lambda\in\mathbb{R}$,
\begin{align}
  \mathbb{E}\! \left[\mathrm{e}^{\lambda\eta^{(t)}}\,\vert\, a^{(1:t)}, \eta^{(1:t-1)}\right] \leq \exp\!\left(\frac{\lambda^2 R^2}{2}\right).
\end{align}
In other words, $\eta^{(t)}$ is conditionally $R$-sub-Gaussian\cite{linear_bandit_1}.
Furthermore, the agent observes the reward of only the chosen arm; thus the rewards of the other arms are not revealed to the agent.

The CMAB problem can be expressed as follows:
\begin{align}
  \label{eq:con_bandit_goal}
  \underset{(a^{(t)})_{t\in\{1,\dots,T\}}}{\text{minimize}} \hspace{5mm} \sum_{t=1}^{T}(r^{(t)}(a^{\star (t)}) - r^{(t)}(a^{(t)})),
\end{align}
where $a^{{\star(t)}}$ is an optimal arm at trial $t$ that satisfies $a^{\star(t)}\coloneqq\argmax_{a^{(t)}\in\mathcal{A}}r^{(t)}(a^{(t)})$.
The objective function $\sum_{t=1}^{T}(r^{(t)}(a^{\star(t)}) - r^{(t)}(a^{(t)}))$ is called the empirical cumulative regret of the agent after $T$ trials \cite{ILOVECONBANDIT}.
To determine the optimal solution of \eqref{eq:con_bandit_goal}, we must know $a^{\star(t)}$ or $\bm{\theta}^{\star}$ in advance; that is, as long as the reward of only the chosen arm is revealed,
it is virtually impossible to solve \eqref{eq:con_bandit_goal}.
Therefore, the CMAB problem is aimed at reducing the number of exploitations to the lowest possible value to rapidly identify the optimal policy without prior information other than the context.

\subsection{LinUCB}
\label{subsec:linucb}
LinUCB algorithms \cite{linucb1,linucb2} are well-known algorithms for solving the linear CMAB problem.
Generally, LinUCB always selects the channel with the highest upper confidence bound to predict of the expected reward.
We refer to LinUCB, which shares coefficient vectors with all arms as JointLinUCB (JLinUCB).
The upper confidence bound of JLinUCB is derived as follows, with reference to \cite{linucb1, linucb2}.
The following inequality holds between the estimated value of the expected reward and its true value:
\begin{align}
  \label{eq:lin_ucb_inequality}
  &\left|\langle\bm{\varphi}(\bm{x}^{(t)}, a^{(t)}), \hat{\bm{\theta}}^{(t)}\rangle - \mathbb{E}[r^{(t)}(a^{(t)})\,\vert\,\bm{x}^{(t)}, a^{(t)}]\right| \nonumber
  \\
  &~~~~\leq \alpha\sqrt{\bm{\varphi}^{\top}(\bm{x}^{(t)}, a^{(t)})(\langle\bm{D}^{(t)},\bm{D}^{(t)}\rangle + \bm{I}_{d})^{-1}\bm{\varphi}(\bm{x}^{(t)},a^{(t)})},
\end{align}
where $\alpha\in\mathbb{R}_{+}$ is a hyperparameter, $\hat{\bm{\theta}}^{(t)}\in\mathbb{R}^d$ is an estimator of $\bm{\theta}^{\star}\in\mathbb{R}^d$ at each trial $t$,
$\langle\bm{D}^{(t)},\bm{D}^{(t)}\rangle \coloneqq \sum_{{t^{\prime}}=1}^{t}\langle\bm{x}_{a_{t^{\prime}},t^{\prime}},\bm{x}_{a_{t^{\prime}},t^{\prime}}\rangle$,
and $\bm{I}_d$ is the $d\times d$ identity matrix.
Let $\bm{D}^{(t)\top}\bm{D}^{(t)} + \bm{I}_{d}$ and $\sum_{t^{\prime}=1}^{t}\bm{D}^{(t^{\prime})}r^{(t^{\prime})}(a^{(t^{\prime})})$ be denoted by $\bm{A}$ and $\bm{b}$, respectively.
Using the right-hand side of \eqref{eq:lin_ucb_inequality}, the score of arm $a^{(t)}\in\mathcal{A}$ (i.e., upper confidence bound) is defined as
\begin{align}
  S_{a^{(t)}} &\coloneqq \langle\bm{\varphi}(\bm{x}^{(t)},a^{(t)}),\hat{\bm{\theta}^{(t)}}\rangle \nonumber
  \\
  &~~~~~~~~~~~~~~~~ + \alpha\sqrt{\bm{\varphi}^{\top}(\bm{x}^{(t)},a^{(t)})\bm{A}^{-1}\bm{\varphi}(\bm{x}^{(t)},a^{(t)})},
  \label{eq:lin_score}
\end{align}
where the second term in \eqref{eq:lin_score} represents $\alpha$ times the standard deviation of $ \langle\bm{\varphi}(\bm{x}^{(t)},a^{(t)}),\hat{\bm{\theta}^{(t)}}\rangle$.
Algorithm~\ref{alg:jointlinucb} provides a detailed description.

\begin{algorithm}[t]
    \floatname{Algorithm}{Algorithm}
    \caption{JointLinUCB}
    \label{alg:jointlinucb}
    \begin{algorithmic}[1]
      \Require{$\alpha > 0$}\\
      \Initialize{$\bm{A} \leftarrow \bm{I}_d$, $\bm{b} \leftarrow \mathbf{0}_d$}
        \For {$t=1, 2,\dots, T$}
          \State Observe context $\bm{x}^{(t)}$
          \State $\bm{\theta}^{(t)} \leftarrow \bm{A}^{-1}\bm{b}$
          \ForAll{$a\in\mathcal{A}$}
          \State Create feature vector $\bm\varphi(\bm{x}^{(t)}, a)$
            \State Calculate $S_{a}$ in \eqref{eq:lin_score}
          \EndFor
          \State \parbox[t]{313pt}{Choose arm $a^{(t)} = \argmax_{a\in\mathcal{A}}S_{a}$ with ties \\ broken arbitrarily\strut}
          \State Observe reward $r^{(t)}(a^{(t)})$
          \State $\bm{A} \leftarrow \bm{A} + \langle\bm{\varphi}(\bm{x}^{(t)}, a^{(t)}),\bm{\varphi}(\bm{x}^{(t)}, a^{(t)})\rangle$
          \State $\bm{b} \leftarrow \bm{b} + \bm{\varphi}(\bm{x}^{(t)}, a^{(t)})\ r^{(t)}(a^{(t)})$
        \EndFor
    \end{algorithmic}
\end{algorithm}

%
%
\section{System Model and Problem Formulation}
\label{sec:system_model}
This section introduces the system model and the decentralized channel allocation problem in WLANs. 
\subsection{System Model}
\label{subsec:system_model}
It is assumed that there are $K$ APs in a square area and $C$ available orthogonal channels with the same bandwidth.
Let the index set of all APs be denoted by $\mathcal{K} \coloneqq \{1, 2,\dots,K\}$ be the index set of all the available channels by $\mathcal{C} \coloneqq \{1, 2,\dots,C\}$,
and the selected channel of AP~$k\in\mathcal{K}$ by $c_k\in\mathcal{C}$.
We denote the index set of APs that lie within the carrier sensing range of AP~$k$ by $\mathcal{N}_k$.
We refer to AP~$i\in\mathcal{N}_k$ as the neighboring AP of AP~$k$.
We only consider downlink transmission, in which each AP transmits a frame in accordance with the CSMA/CA protocol.

To model the contention relationships among APs, we use a contention graph $\mathcal{G}^{(t)} \coloneqq (\mathcal{K}, \mathcal{E}^{(t)})$ at trial $t$: 
where the nodes represent APs, and the edge set $\mathcal{E}^{(t)}$ is defined by $\mathcal{E}^{(t)} \coloneqq \{\{k, k^{\prime}\}\,\vert\, k\in\mathcal{K}\land k^{\prime}\in\mathcal{N}_k\land c_k^{(t)} = c_{k^{\prime}}^{(t)}\}$, that is,
the edges $e_{k, k^{\prime}}^{(t)} \coloneqq \{k,k^{\prime}\}\in\mathcal{E}^{(t)}$ are connected 
only when AP~$k$ and AP~$k^{\prime}$ are within the carrier sensing range and they select the same channel.

In contrast to previous studies\cite{compare_wlan_2019,compare_wlan_2020}, we assume that the AP~$k$ can obtain the channels selected by the neighboring APs as prior information.
Note that each AP~$k$ does not know other information regarding the neighboring APs (e.g., traffic) and any information about APs other than the neighboring APs.
Furthermore, we assume that no information is available on the throughput in advance, and the throughput observed by AP~$k$ follows a probability distribution.

\subsection{Decentralized Channel Allocation Problem in WLANs}
\label{subsec:channel_allocation}
We formulate a decentralized channel allocation problem in an unknown environment 
in which the access probability of each AP and the throughput model are not known in advance.

We first define $p_k$ as the transmission probability of AP~$k$ as follows.
Let $T_{\mathrm{slots}}$ be a period in which AP~$k$ is either always attempting to transmit with probability $p_k\in[0,1]$
or not attempting to transmit at all with probability $1-p_k$, where the probability $p_k$ is time-invariant.
For a sufficiently long period, the sum of the actual frame transmission times is proportional to $p_k$.
The value of $p_k$ is considered to be a measure of the access probability of AP~$k$.

The objective of this channel allocation problem is to maximize the sum of the system throughput of each channel allocation experienced by learning.
We let $R^{(t)}(\mathcal{K}, \mathcal{C})$ denote the system throughput.
In this study, our optimization problem is formulated as follows:
\begin{align}
  \underset{(c_k^{(t)})_{t\in\{1,\dots,T\}, k\in\{1,\dots,K\}}}{\text{maximize}} \hspace{5mm} \sum_{t = 1}^{T} R^{(t)}(\mathcal{K}, \mathcal{C}), 
\end{align}
where $R^{(t)}(\mathcal{K}, \mathcal{C}) \coloneqq \sum_{k\in\mathcal{K}}f_k(c_k^{(t)}, \bm{c}^{(t)}_{\mathcal{N}_k}, \bm{p}_{\mathcal{N}_k})$.
In the above problem, $\bm{c}_{\mathcal{N}_k}$ denotes the vector of the channels of the neighboring APs of AP~$k$, and $\bm{p}_{\mathcal{N}_k}$ denotes the vector of the transmission probabilities of the neighboring APs of AP~$k$.
Note that for AP~$k$, the values of $p_{k^{\prime}}$ ($k^{\prime}\in\mathcal{K}\setminus k$) are unknown.
The function $f_k(c_k^{(t)}, \bm{c}^{(t)}_{\mathcal{N}_k}, \bm{p}_{\mathcal{N}_k})$ is treated as the throughput for convenience.
However, in the following discussion, any function may be used
as long as $f_k(c_k^{(t)}, \bm{c}^{(t)}_{\mathcal{N}_k}, \bm{p}_{\mathcal{N}_k})$ is an evaluation measure based on the channels and access probabilities.

\section{Proposed Scheme}
\label{sec:propose}
To achieve this objective, we implement an MAB-based scheme in which traffic condition-wise channel allocation is learned by all APs in a distributed manner without sharing information.
\subsection{Overview of Proposed Scheme}
\label{subsec:overview}
An overview of the proposed method is shown in Fig.~\ref{fig:proposed_scheme} in Section~\ref{sec:introduction}.
Each AP~$k\in\mathcal{K}$ selects a channel for its own environment by leveraging the channel set of neighboring APs~$N_k$ as prior information.
Each AP has a different environment and does not share learned information among APs.
By repeating the procedures of steps 1--7 in Fig.~\ref{fig:proposed_scheme}, 
we obtain a channel allocation strategy that improves the system throughput, as defined in the channel allocation problem.

This sequence of trials (i.e., steps 1--6 in Fig.~\ref{fig:proposed_scheme}) can be formulated as a CMAB problem.
However, the following challenges must be addressed:
\begin{itemize}
  \item The resultant performance of the MAB approach is strongly affected by how to incorporate prior information into training.
  Although, as mentioned in Remark~1, we need to design the feature vectors with reference to the context so that the APs can learn the allocations properly, 
  it is unclear how to construct them in this problem.
  \item Selfishly performing CMAB learning in a multi-agent environment leads to cycles specific to the channel allocation problem 
  because the CMAB algorithm performs either exploitation or exploration. Exploitation can be interpreted as taking actions that maximize one's own benefit only, and exploration as taking a suboptimal action at the moment.
  When the channel fluctuation is severe, the WLAN system will be heavily loaded.
\end{itemize}

As a solution to the former, we propose CDFE in Section~\ref{subsec:feature}, which focuses on the shape of competing graphs.
For the latter, in Section~\ref{subsec:p_lin}, we extend JLinUCB to restrict the exploratory action by discounting rewards.
We call it Penalized JLinUCB (P-JLinUCB).

\subsection{Contextual Multi-Armed Bandit Formulation}
\label{subsec:bandit_fomulation}
In this section, we formulate the channel selection problem as a CMAB problem.
Consider AP~$k$ as a leaning agent. AP~$k$ repeatedly observes a context, selects an arm, and observes a reward per $T_{\mathrm{slots}}$.
Since AP~$k$ can know the channel set of neighboring APs as prior information, 
we let $\bm{c}^{(t)}_{\mathcal{N}_k}$ be the context vector of AP~$k$.
The design of feature vectors using $\bm{c}^{(t)}_{\mathcal{N}_k}$ is described in detail in the following section.
In this problem, as the AP selects the channel in each trial, the channel set $\mathcal{C}$ is the arm set.

The objective of AP~$k$, as expressed in \eqref{eq:con_bandit_goal}, can be rewritten as follows:
\begin{align}
  \label{eq:goal}
  \underset{(c_{k}^{(t)})_{t\in\{1,\dots,T\}}}{\text{minimize}} \hspace{5mm} \sum_{t=1}^{T}\!\left(r^{(t)}(c^{\star(t)}_{k}) - r^{(t)}(c^{(t)}_{k})\right), 
\end{align}
where $c_{k}^{(t)}$ denotes the channel selected by AP~$k$ at trial $t$.
As mentioned in Section~\ref{subsec:cmab}, $c^{\star(t)}_{k}$ is not known in advance; therefore AP~$k$ needs to appropriately exploit and explore.
Furthermore, in a real environment, the access probabilities of neighboring APs are assumed to fluctuate over time.
Hence, AP~$k$ must learn the optimal channel as quickly as possible.
From the two requirements mentioned above, we need to properly construct a feature vector.

\subsection{Contention-Driven Feature Extraction}
\label{subsec:feature}
In this study, because $\bm{c}^{(t)}_{\mathcal{N}_k}$ (i.e., the channel set of neighboring APs) is utilized in advance, 
we can first naturally construct the feature vector as follows:
\begin{align}
  \label{eq:simple_feature}
  \bm{\varphi}_1{(\bm{c}^{(t)}_{\mathcal{N}_k},c_k^{(t)})} \coloneqq (c^{(t)}_k, \bm{c}^{(t)}_{\mathcal{N}_k})^{\top},
\end{align}
where $\bm{c}^{(t)}_{\mathcal{N}_k}$ denotes the channel set of neighboring APs at trial $t$.
In this case, the number of features is $C^{|\mathcal{N}_k|+1}$, where $|\mathcal{N}_k|$ denotes the number of elements (i.e., cardinality) of the set $\mathcal{N}_k$.
This depends on both the number of channels and the number of APs, which is undesirable in terms of learning efficiency.

Second, to reduce the number of features, we identify the channel set of neighboring APs 
that can be considered identical for learning.
This process is referred to as CDFE.
\begin{figure}[!t]
  \centering
  \includegraphics[width=\columnwidth]{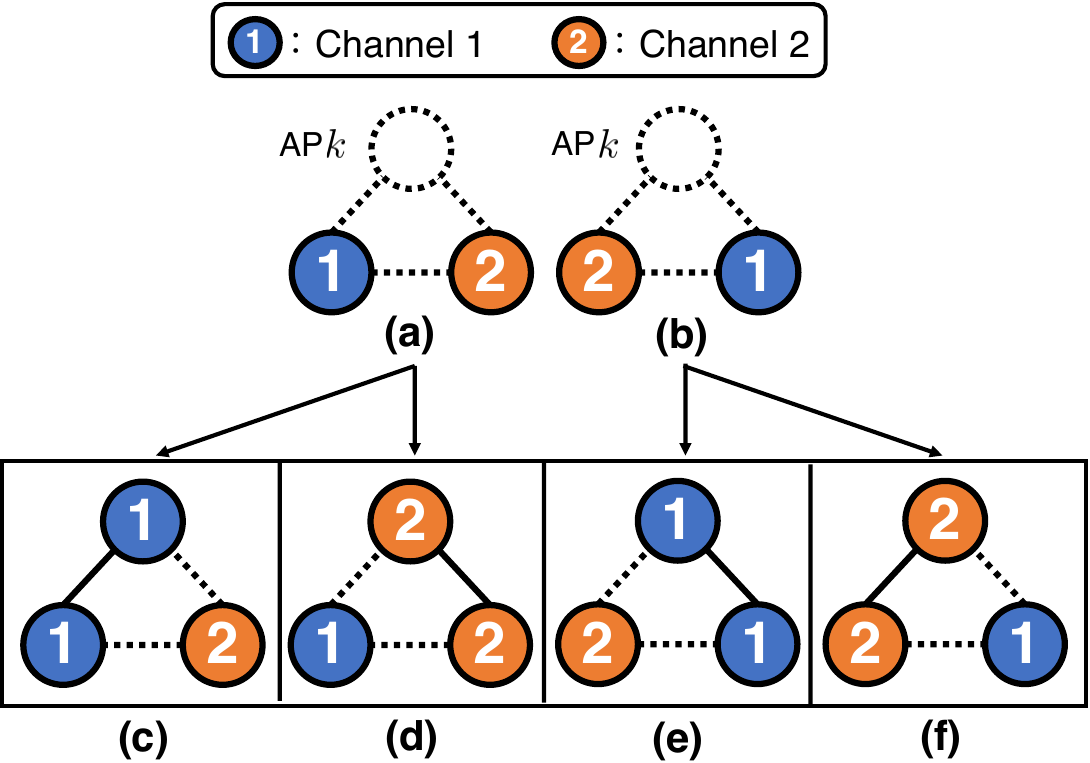}
  \caption{For two different channel sets observed by AP~$k$ (a), (b), there are four possible channel allocations: (c), (d), (e) and (f) (the number of AP $K=3$, and available channels $C=2$). 
Note that only two types of contention graphs exist.}
\label{fig:contention_graph}
\end{figure}
The CDFE is based on the idea that the distribution of the throughput changes depending on the form of the contention graph.
Using CDFE, we can organize information as in the following example.
Fig.~\ref{fig:contention_graph} presents two environments (a) and (b) where, in the case $K=3, C=2$, only the context $\bm{c}^{(t)}_{\mathcal{N}_k}$ is different from each other, and four possible channel allocations (c), (d), (e), and (f).
While the pairs ((c), (f)) or ((d), (e)) have different channel allocations, they have the same environment in terms of the reward generation process.
This fact suggests that the number of environments to be learned can be reduced by classifying contexts in the form of a contention graph. 

The feature vector with CDFE is defined as follows:
\begin{align}
  \label{eq:feature}
  &\bm{\varphi}_2{(\bm{c}^{(t)}_{\mathcal{N}_k},c_k^{(t)})} \coloneqq \left(1, \phi_1,\dots, \phi_{|\mathcal{N}_k|}\right)^{\top},\\
  &\phi_i \coloneqq
  \begin{cases}
    1 & \text{if}\ e^{(t)}_{k,i}\in\mathcal{E}^{(t)} \\
    0 & \text{otherwise}
  \end{cases},\
  i \in \mathcal{N}_k.
  \label{eq:feature_factor}
\end{align}
For each channel that AP~$k$ can select, the feature vector indicates which neighboring AP occupies that channel at trial $t$ (i.e., this feature is a vector representation of the contention graph around the target AP).
The total number of features with extraction equals $2^{|\mathcal{N}_k|}$, and it does not depend on the number of available channels $C$.
Since the base of the index is fixed at 2, the increase in the number of features with respect to the number of APs is more gradual than that of the features in \eqref{eq:simple_feature}.
Furthermore, it is worth noting that, since the reward is modeled by \eqref{eq:linear_reward}, 
by updating $\bm{\theta}$ via the feature vector with CDFE, each element of $\bm{\theta}$ corresponds to a measure of the impact of each AP on the reward. 
This implies that APs can disentangle a channel utility function (i.e., reward) with respect to contending neighboring APs and that using \eqref{eq:cmab_reward}, the expected reward for each channel is accurately estimated regardless of the channel variation of neighboring APs.

\subsection{Penalized JointLinUCB}
\label{subsec:p_lin}
When each AP is trained independently using the CMAB algorithm, channel allocation is not always converged.
This is because each AP focuses on exploitation as the CMAB learning progresses and therefore the optimal actions of the APs vary at each trial in our system model 
where the environment changes depending on the actions of neighboring APs.

\begin{figure}[t!]
  \centering
  \includegraphics[width=\columnwidth]{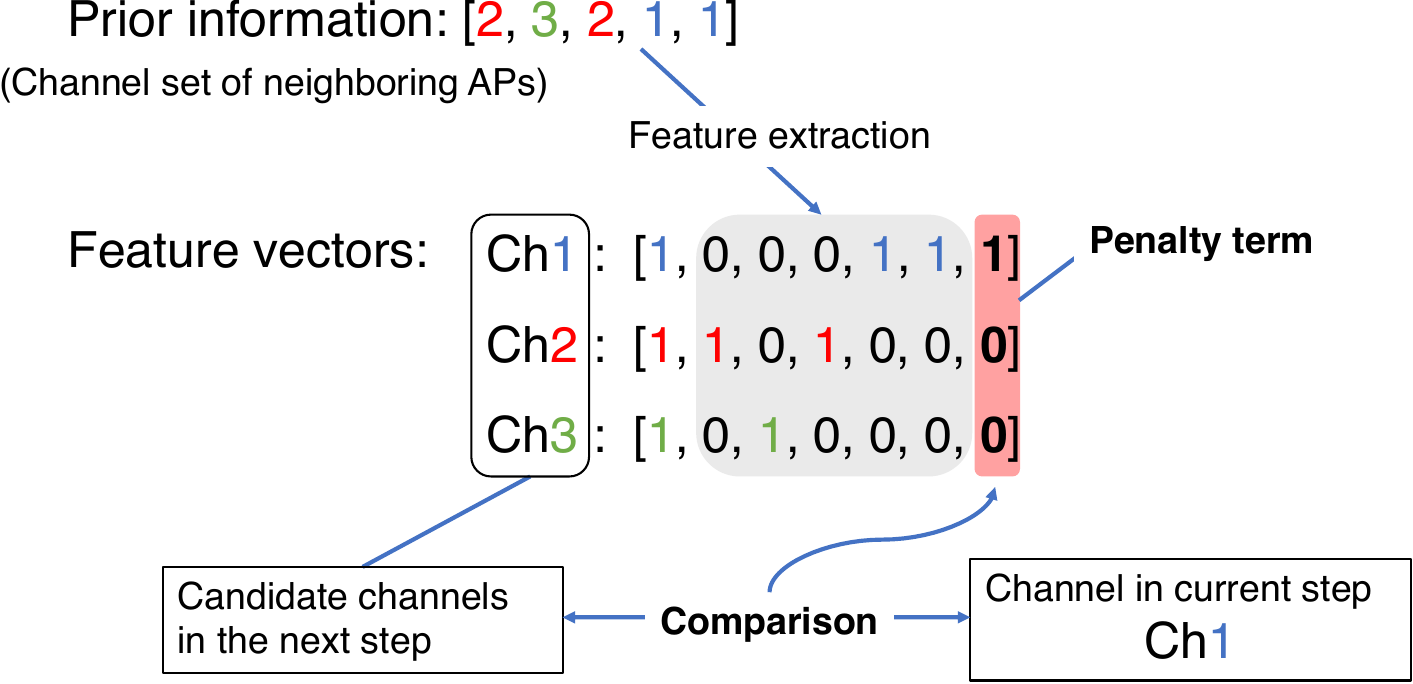}
  \caption{Example of feature construction for Penalized JLinUCB (the number of APs $K=6$ and available channels $C=3$).}
  \label{fig:penalty_feature}
\end{figure}
  
\begin{algorithm}[t]
  \floatname{Algorithm}{Algorithm}
  \caption{Penalized JointLinUCB (P-JLinUCB)}
  \label{alg:plin}
  \begin{algorithmic}[1]
    \Require{$\alpha$, $\beta$, $\bm{A}$, $\bm{b}$}
    \Ensure{$\bm{\theta}$, $\bm{A}$, $\bm{b}$}
      \State $\bm{\theta} \leftarrow \bm{A}^{-1}\bm{b}$.
      \State Observe context $\bm{x}^{(t)}$
        \ForAll{$c\in\mathcal{C}$}
        \State Create feature vector $\bm\varphi(\bm{x}^{(t)}, c)$
        \State Add a penalty term to the feature vector
          \State Calculate $S_{c}$ in \eqref{eq:lin_score}.
        \EndFor
        \State \parbox[t]{313pt}{Choose a channel $c_{k}^{(t)} = \argmax_{c\in\mathcal{C}}S_{c}$ with ties \\ broken arbitrarily\strut}
        \State Observe reward $r^{(t)}(c_{k}^{(t)})$
        \If{$c_{k}^{(t)} \neq c_{k}^{(t-1)}$}
        \State $r^{(t)}(c_{k}^{(t)}) \leftarrow \beta r^{(t)}(c_{k}^{(t)})$ 
        \EndIf
        \State $\bm{A} \leftarrow \bm{A} + \langle\bm\varphi(\bm{x}^{(t)}, c_k^{(t)}),\bm\varphi(\bm{x}^{(t)}, c_k^{(t)})\rangle$
        \State $\bm{b} \leftarrow \bm{b} + \bm\varphi(\bm{x}^{(t)}, c_k^{(t)})\ r^{(t)}(c_{k}^{(t)})$
        \State\Return{($\bm{\theta}$, $\bm{A}$, $\bm{b}$)}
  \end{algorithmic}
\end{algorithm}

To address this challenge, we propose the P-JLinUCB algorithm, which is an extension of JLinUCB\cite{linucb1,linucb2}.
Algorithm~\ref{alg:plin} provides its detailed description.
The key steps of this algorithm are \textit{1)} adopting the parameter to discount the observed rewards and \textit{2)} building a feature vector in the linear model so that the penalties are incorporated into JLinUCB.
In detail, this algorithm updates the parameters of the JLinUCB by discounting the observed reward by $\beta\in[0,1]$ 
when the channel to be selected as a result of the AP computing $S_c$ is different from the current channel.
However, if we simply discount the reward, each AP cannot associate the reason for the discount with the channel changes based only on the current channels of the neighboring APs.
Hence, as context information, we introduce an additional index, which is an indicator of whether or not the channel has changed into the feature vector.

An example of the reconstruction of the feature vector with extraction is shown in Fig.~\ref{fig:penalty_feature}.
Among the feature vectors subjected to CDFE described in Section~\ref{subsec:feature}, 
1 is added at the end of the feature vector corresponding to the same channel as the current one; otherwise, 0 is added as an element.
In summary, the product of the term at the end of the feature vector and the element of $\bm{\theta}$ functions as a penalty term.

\section{Numerical Evaluation}
\label{sec:evaluate}
\subsection{Setup}
\label{subsec:setup}
We assume a WLAN system with 10 APs in a 1000\,m $\times$ 1000\,m area, that is, $K=10$. 
The carrier sensing range of the AP is a circle with a radius of 550\,m centered on the AP\cite{csrange1,csrange2,nakashima}, and the number of available channels $C$ is three, which is equal for all APs.
The total number of learning trials $T$ is set to 10,000.
Since $K=10$, out of 10,000 trials, each AP performs CMAB learning only in 1,000 trials.
The reward $r^{(t)}(c_{k}^{(t)})$ is defined as follows:
\begin{align}
  r^{(t)}(c_{k}^{(t)}) \coloneqq \frac{1}{1+\sum_{i\in\mathcal{N}_k}X_{p_i}\cdot \ind(c_{k}^{(t)}=c_{i}^{(t)})}.
  \label{eq:reward}
\end{align}
where $X_{z}\sim\mathrm{Ber}(z)$, which is a random variable that follows a Bernoulli distribution with an expected value $z$.
Under the assumption described in Section~\ref{subsec:channel_allocation}, 
the reward can be regarded as the ratio of the transmission time AP~$k$ acquired during $T_{\mathrm{slots}}$.
In this numerical evaluation, $r^{(t)}(c_{k}^{(t)})$ corresponds to $f_k(c_k^{(t)}, \bm{c}^{(t)}_{\mathcal{N}_k}, \bm{p}_{\mathcal{N}_k})$.

To investigate whether the proposed method can obtain a channel allocation strategy based on the traffic conditions of neighboring APs, we set the transmission probability of AP~$k\in\mathcal{K}$ (i.e., $p_k$) to be uniformly random.
We also evaluate the case where the transmission probabilities of all APs are identical, 0.5, as the baseline.
As described in Section~\ref{subsec:system_model}, each AP is assumed to have no prior knowledge of the transmission probabilities of the other APs.

\subsection{Evaluation of Channel Exploration Performance}
\label{subsec:exploration_performance} 
\begin{figure*}[t!]
  \centering
  \subfigure[Estimated reward of ch~1.]{
    \includegraphics[width=0.3\linewidth]{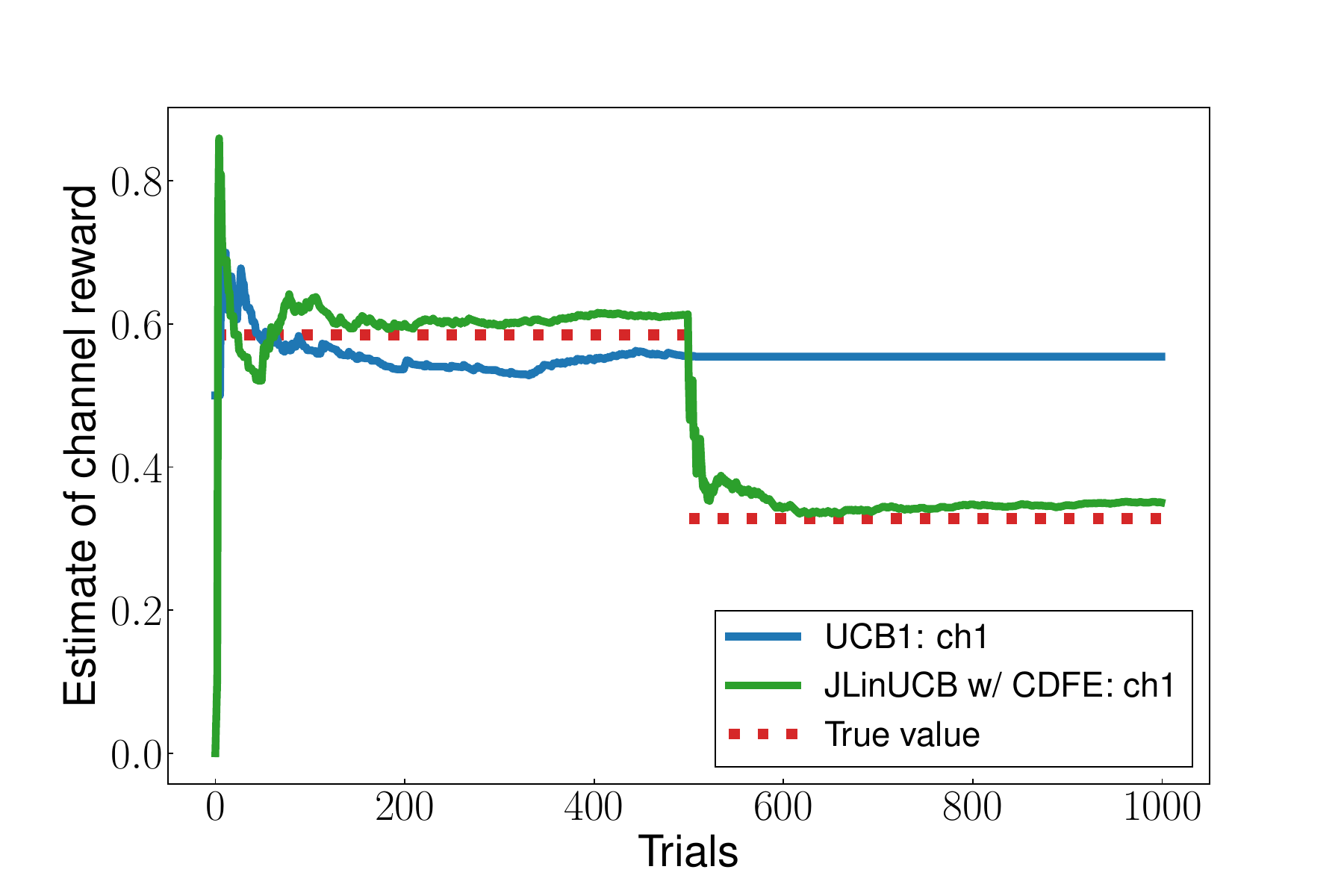}
    \label{fig:ch1}
  }
  \subfigure[Estimated reward of ch~2.]{
    \includegraphics[width=0.3\linewidth]{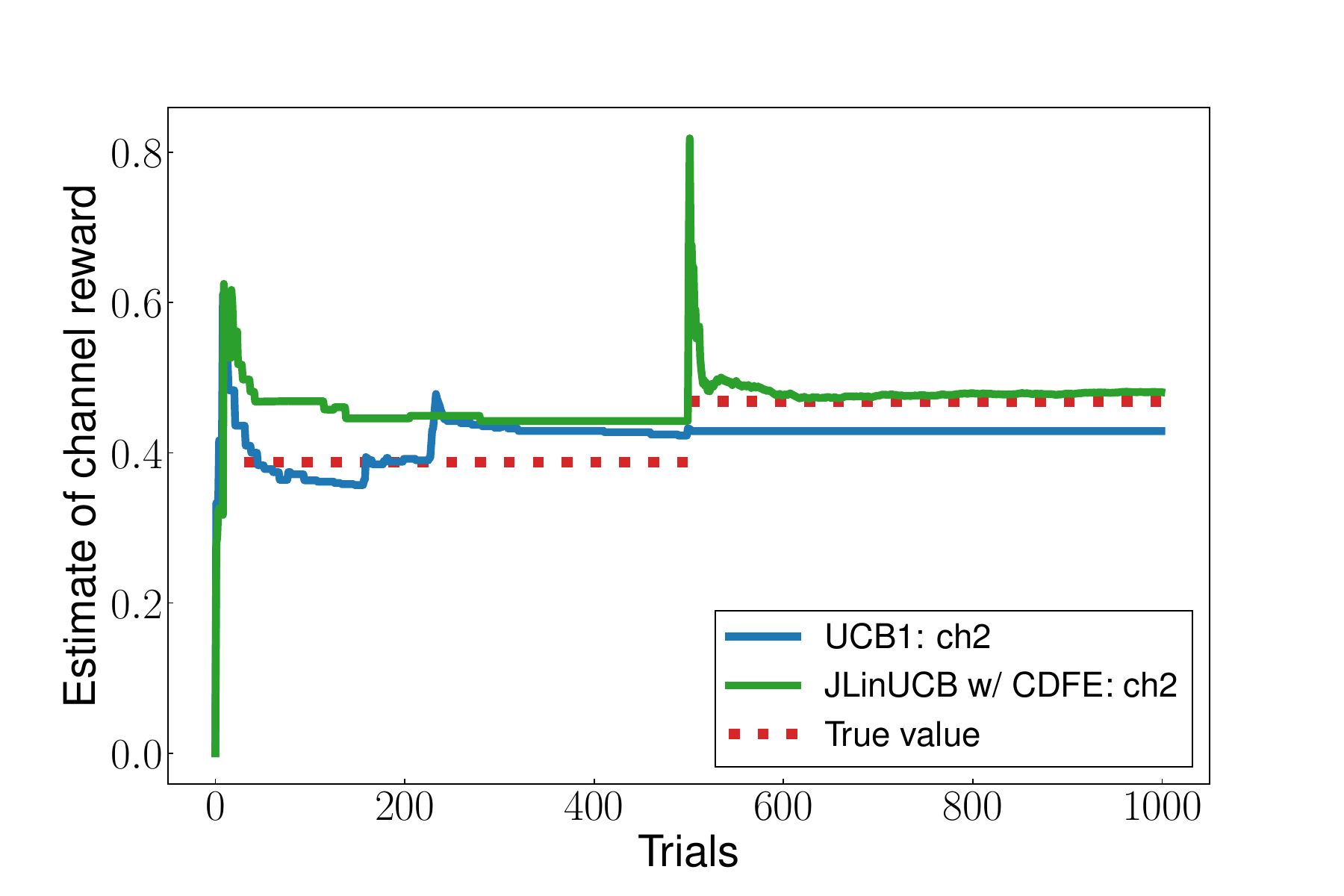}
    \label{fig:ch2}
  }
  \subfigure[Estimated reward of ch~3.]{
    \includegraphics[width=0.3\linewidth]{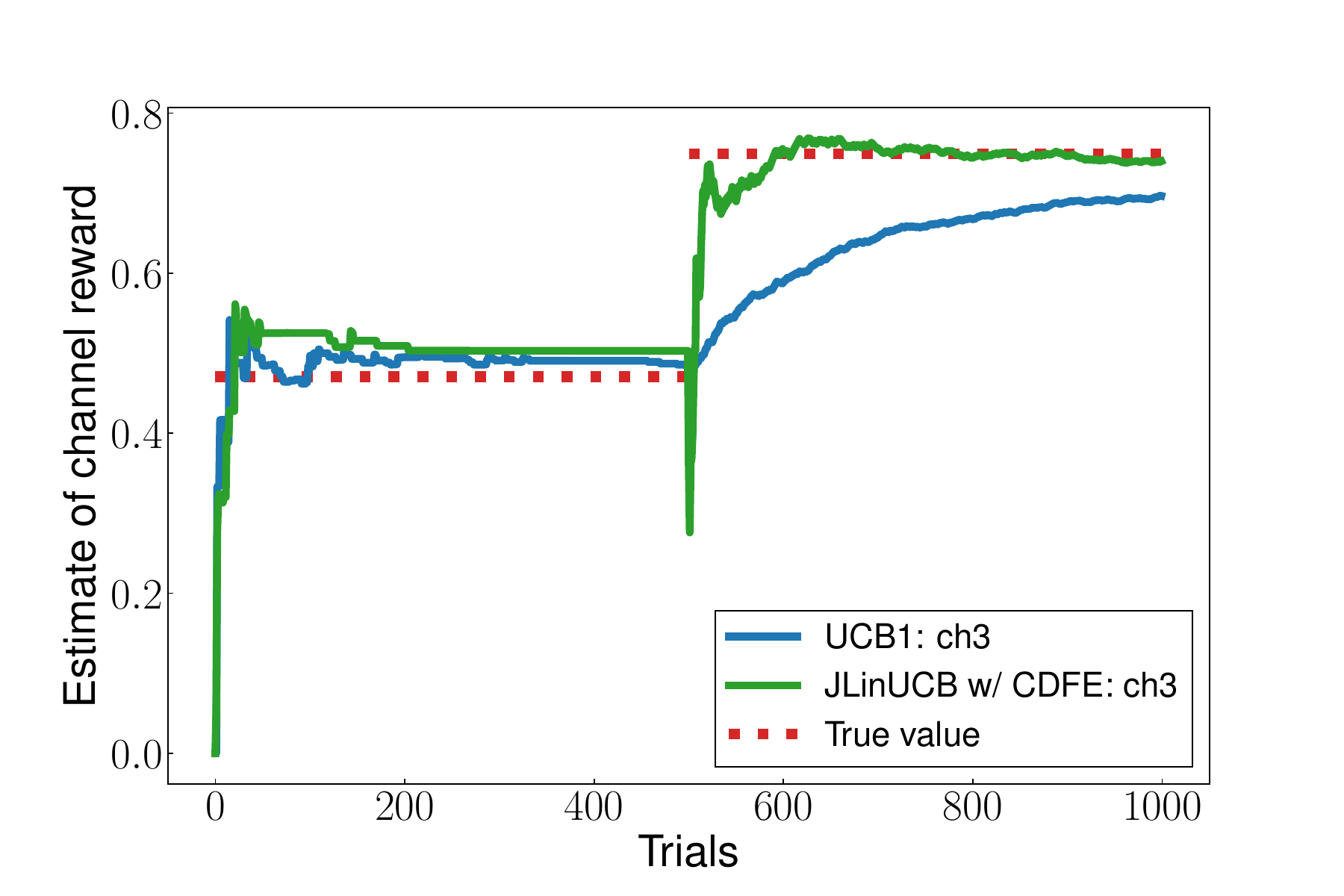}
    \label{fig:ch2}
  }
  \caption{Estimate of the reward for each channel learned by one AP. 
  Note that 9 neighboring APs adjust their channels in just 500 trials.
}
\label{fig:ave_prob}
\end{figure*}

In this section, we focus on a single learning AP and compare the proposed method with simple MAB learning in terms of the estimated channel rewards.
Specifically, we compare the following two methods:
\begin{itemize}
  \item UCB1\cite{ucb}, which is one of the well-known MAB algorithms. Note that this does not leverage any prior information and corresponds to the method of existing studies\cite{compare_wlan_2019,compare_wlan_2020}.
  \item JLinUCB using extracted features $\bm{\varphi}_2$ in \eqref{eq:feature}, which is referred to as ``JLinUCB w/ CDFE''
\end{itemize}
The values of the hyperparameter $\alpha$ is set to 0.8.
For simplicity, we assume that all nine APs around one AP in focus are within the carrier sense range, 
and the transmission probabilities of them are identical, 0.5.

To confirm that the proposed method can follow the variation of the true channel reward, we let the neighboring APs adjust their channels only on exactly the 500th trial. 
In particular, the selected channels of AP~2--AP~10 are adjusted from $(2, 2, 2, 2, 3, 3, 3, 1, 1)$ to $(1, 1, 1, 1, 1, 3, 2, 2, 2)$.
In other words, the channel of the neighboring AP did not change for the first 499 trials and for the last 500 trials.

\begin{table}[t]
  \centering
  \caption{Number of channel explorations before and after 500th trial.}
  \begin{tabular}{cccc}
    \toprule
    \multirow{2}{*}{Method}&\multirow{2}{*}{Channel}&\multicolumn{2}{c}{Trials} \\
    &&1--499&501--1000\\
    \midrule
    \multirow{3}{*}{UCB1}&
    Ch~1 & 328& 4\\
    &Ch~2 & 59& 4\\
    &Ch~3 & 112& 492\\
    \midrule
    \multirow{3}{*}{JLinUCB w/ CDFE}&
    Ch~1 & 452& 2\\
    &Ch~2 & 16& 5\\
    &Ch~3 & 31& 493\\
  \bottomrule
  \end{tabular}
  \label{tab:true_reward}
\end{table}

Fig.~\ref{fig:ave_prob} shows the transition of the AP's estimated channel reward value for each trial, and Table~\ref{tab:true_reward} lists the number of explorations for each channel.
In the first half of the trials (i.e., 1--499 trials), both methods almost correctly explore, i.e., estimate the channel reward. 
In contrast, in the second half, the differences between the methods are quite visible.
While the proposed method immediately responds to fluctuations in all true channel rewards and estimates them correctly with a few explorations, 
UCB1 gradually follows the 
fluctuations in rewards by intensively exploring channel 1 again.
Concomitantly, UCB1 explores the channel 2 and 3 for a few times, so the estimated value has not changed from the first half.
This is because of the lack of prior information about neighboring APs under contention and the entangled form of reward function. 
Namely, in UCB1, the AP of interest could not detect neighboring APs varying the channel nor re-assess the reward function in view of the channel variation.
This comparison indicates the effectiveness of the proposed method possessing the disentangled form of the reward and reforming the reward function according to the channel variation of neighboring APs.
In summary, the proposed method with prior information and AP-wise disentanglement of rewards outperforms the existing studies\cite{compare_wlan_2019,compare_wlan_2020} in terms of the exploration performance.

\subsection{Evaluation of Channel Allocation Performance}
\label{subsec:allocation_performance} 
We compared the average system throughput, represented by $R^{(t)}(\mathcal{K},\mathcal{C})$, 
in 10 different topologies with APs randomly placed in a square area using the following five methods:
\begin{itemize}
  \item UCB1\cite{ucb}, which is one of the well-known MAB algorithms. Note that this does not leverage any prior information and corresponds to the method of existing studies\cite{compare_wlan_2019,compare_wlan_2020}.
  \item JLinUCB using features $\bm{\varphi}_1$ in \eqref{eq:simple_feature}, which is referred to as ``JLinUCB w/o CDFE''
  \item JLinUCB using extracted features $\bm{\varphi}_2$ in \eqref{eq:feature}, which is referred to as ``JLinUCB w/ CDFE''
  \item P-JLinUCB using features $\bm{\varphi}_1$ in \eqref{eq:simple_feature}, which is referred to as ``P-JLinUCB w/o CDFE''
  \item P-JLinUCB using extracted features $\bm{\varphi}_2$ in \eqref{eq:feature}, which is referred to as ``P-JLinUCB w/ CDFE''
\end{itemize}
The values of the hyperparameter $\alpha$ and reward discount parameter $\beta$ were both set to 0.8.

\begin{figure}[t!]
  \centering
  \subfigure[Identical transmission probability for all APs ($p_k$ is set to identical 0.5 for all $k\in\mathcal{K}$).]{
    \includegraphics[width=0.9\columnwidth]{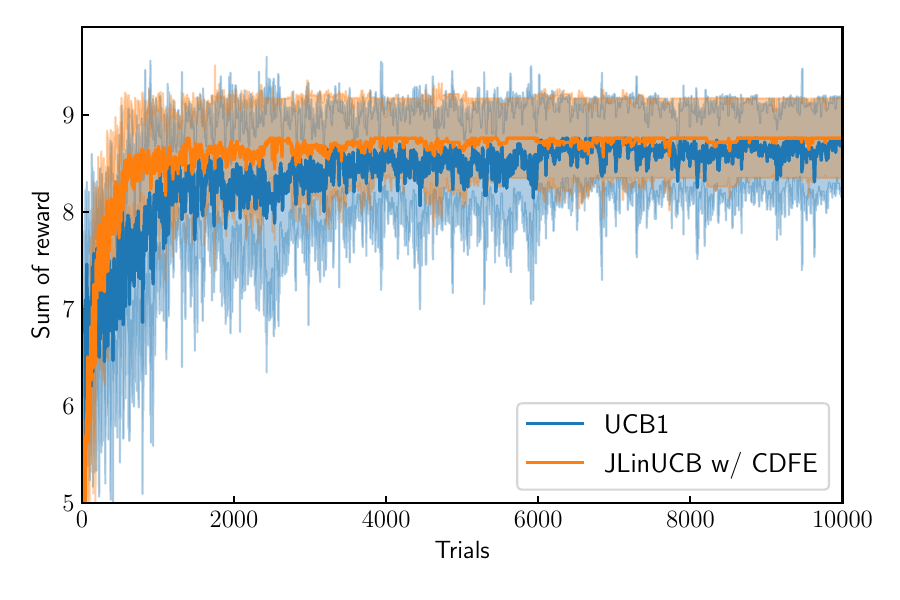}
    \label{fig:ucb_same}
  }
  \subfigure[Nonidentical transmission probability for all APs ($p_k$ is set uniformly at random for all $k\in\mathcal{K}$).]{
    \includegraphics[width=0.9\columnwidth]{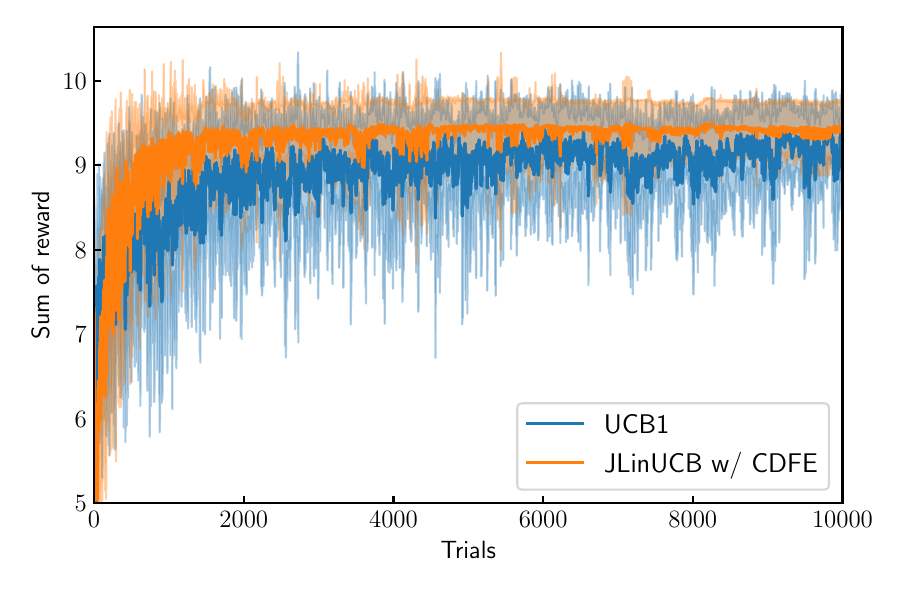}
    \label{fig:ucb_dif}
  }
  \caption{Comparison of MAB-based channel allocation schemes with and without prior information (i.e., JLinUCB w/ CDFE and UCB1, respectively), by average total reward, i.e., system throughput, over 10 random topologies of APs.
  Shaded regions denote the standard deviation of the performance.}
  \label{fig:ucb}
\end{figure}
\begin{figure}[t!]
  \centering
  \subfigure[Identical transmission probability for all APs ($p_k$ is set to identical 0.5 for all $k\in\mathcal{K}$).]{
    \includegraphics[width=0.9\columnwidth]{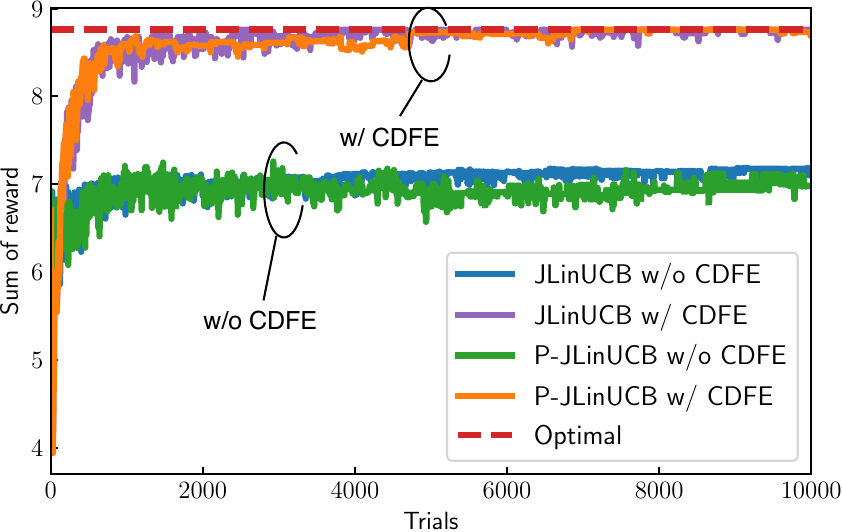}
    \label{fig:prod_same}
  }
  \subfigure[Nonidentical transmission probability for all APs ($p_k$ is set uniformly at random for all $k\in\mathcal{K}$).]{
    \includegraphics[width=0.9\columnwidth]{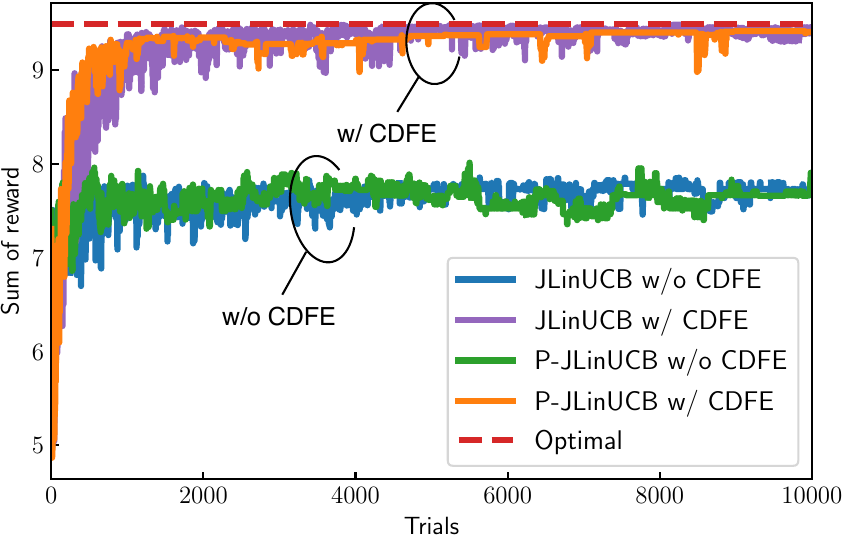}
    \label{fig:prod_dif}
  }
  \caption{Comparison of P-JLinUCB/JLinUCB-based channel allocation schemes using feature vector in \eqref{eq:simple_feature} and using feature vector in \eqref{eq:simple_feature} (i.e., w/o CDFE and w/ CDFE, respectively) by average total reward, i.e., system throughput, over 10 random topologies of APs.
  }
  \label{fig:prod}
\end{figure}

\subsubsection{Effect of Using Prior Information}
First, in terms of the system throughput, we evaluated the effectiveness of utilizing the channels of neighboring APs as prior information for WLAN channel allocation.
Fig.~\ref{fig:ucb} compares the results of channel allocation based on UCB1, which leverages no prior information, and JLinUCB with CDFE, which leverages prior information, using system throughput as a measure of performance. 
The sum of the rewards for identical traffic conditions is shown in Fig.~\ref{fig:ucb}\subref{fig:ucb_same}, and that for nonidentical traffic conditions is shown in Fig.~\ref{fig:ucb}\subref{fig:ucb_dif}.
In both cases, we observed that JLinUCB with CDFE outperformed UCB1 overall in terms of system throughput. 
We also found that JLinUCB with CDFE had a smaller variance in system throughput than UCB1.
These results indicate that the proper use of the channel of neighboring APs as prior information improves system throughput and stability.
Furthermore, it is worth mentioning that JLinUCB maintains its performance regardless of whether the traffic conditions are identical or nonidentical and enables the learning of traffic condition-wise channel allocation.

\begin{table*}[t]
  \centering
  \caption{Average number of channel adjustments per 2,000 trials.}
  \begin{tabular}{ccccccc}
    \toprule
    \multirow{2}{*}{Transmission probability}&\multirow{2}{*}{Method}&&&Trials&& \\
    &&1--2000&2001--4000&4001--6000&6001--8000&8001--10000\\
    \midrule
    \multirow{3}{*}{Identical (Fig.~\ref{fig:ucb}\subref{fig:ucb_same} and Fig.~\ref{fig:prod}\subref{fig:prod_same})}&\bf{P-JLinUCB w/ CDFE}&\bf{109.1}&\bf{7.6}&\bf{8.8}&\bf{5.0}&\bf{2.1}\\
    &JLinUCB w/ CDFE&505.3&21.8&144.7&139.6&147.2\\
    &UCB1&621.3&356.7&278.3&184&179.7\\
    \midrule
    \multirow{3}{*}{Nonidentical (Fig.~\ref{fig:ucb}\subref{fig:ucb_dif} and Fig.~\ref{fig:prod}\subref{fig:prod_dif})}&\bf{P-JLinUCB w/ CDFE}&\bf{96.4}&\bf{5.6}&\bf{0.5}&\bf{2.1}&\bf{0.9}\\
    &JLinUCB w/ CDFE&813&292.5&207.6&211&145.3\\
    &UCB1&819&507&435&415&364\\
  \bottomrule
  \end{tabular}
  \label{tab:system_regret}
\end{table*}
\subsubsection{Validity of Contention-Driven Feature Extraction.}
To confirm the validity of the proposed CDFE-based feature vector design, we conducted CMAB learning using the feature vectors defined in \eqref{eq:simple_feature} and \eqref{eq:feature}.
Fig.~\ref{fig:prod} shows the learning results of JLinUCB/P-JLinUCB with and without CDFE by the transition of the total reward of all APs during 10,000 learning trials.
For reference, the result of the centralized control of channel allocation with a known contention graph $\mathcal{G}^{(t)}$ and transmission probabilities for all APs is illustrated as ``Optimal''.
Similar to Fig.~\ref{fig:ucb}, Fig.~\ref{fig:prod}\subref{fig:prod_same} shows the results when the transmission probabilities of all the APs are set to an identical value of 0.5,
whereas, Fig.~\ref{fig:prod}\subref{fig:prod_dif} shows the results when the transmission probabilities are set uniformly at random.
In both cases, we can see that the difference between the system throughput obtained by the approach using CDFE and the optimal value is quite small.
However, the approach using simple feature vectors $\phi_1$ in \eqref{eq:simple_feature} causes a significant degradation in the system throughput from the optimal value.
This indicates that for the channel allocation problem, the designed features based on contention graphs as a linear model 
in decentralized learning are effective in increasing the system throughput regardless of traffic conditions.

\subsubsection{Performance Evaluation of Penalized JointLinUCB}
In Fig.~\ref{fig:prod}, we also confirm that the fluctuation in the sum of rewards per trial is smaller for P-JLinUCB with CDFE when compared with that of LinUCB with CDFE. 
This is attributed to the number of channel adjustments.
The number of channel adjustments is an important index because the burden on the system becomes enormous when the channel fluctuation during learning is significant.
Table~\ref{tab:system_regret} lists the average number of channel adjustments per 2,000 trials.
We can see that the number of channel adjustments is significantly reduced using P-JLinUCB.
On average, there is no major difference in the system throughput, as shown in Fig.~\ref{fig:prod}.  
This indicates that the number of channel adjustments can be suppressed without degrading the performance by introducing penalties. 
As discussed in Section~\ref{subsec:feature}, these results can be attributed to the fact that the same reward can be expected when the environment is isomorphic as a contention graph, 
regardless of the channel set of neighboring APs; that is, the number of channel allocation patterns to be explored, is diminished by CDFE. 
Hence, the expected reward can be predicted even for the channel set that has not been experienced. 
Consequently, CMAB learning performed well, even when the channel adjustment is suppressed to some extent.

Furthermore, using P-JLinUCB with CDFE is expected to obtain a channel allocation with high performance in terms of system throughput, even when the learning is stopped at an arbitrary time. 
This suggests that it is possible not only to reduce the learning cost of AP, but also to relearn instantly when the environment changes.
In contrast, in LinUCB, because APs adjust their channels frequently, the channel allocation performance cannot be guaranteed after an interruption of learning.

\section{Conclusion}
\label{sec:conclusion}
In this study, we proposed a CMAB-based decentralized WLAN channel exploration framework. 
This framework enabled an AP to robustly evaluate the channel against the variation of the reward caused by the neighboring APs varying the channel. 
This framework also improved the convergence probability of channel allocation as a system by suppressing channel adjustment. 
The key idea of the former is disentangling a channel utility function (i.e., reward) with respect to contending neighboring APs. 
This idea is realized by CDFE, which exploits the channels selected by neighboring APs and extracts features corresponding to the adjacencies in the contention graph.
The specific mechanism for the latter is to adjust the reward in JLinUCB with a discount parameter and add a penalty term to the feature vector to model the impact of the discounted reward, which we named P-JLinUCB. 
Numerical evaluations confirmed that the proposed method accurately estimated the updated channel reward with significantly fewer explorations than that of the existing method (i.e., simple MAB), and suppressed the variation in channel allocation.

\section*{Acknowledgment}
This work was supported in part by JSPS KAKENHI Grant Number JP18H01442.

\bibliographystyle{ieicetr}
\bibliography{main}

\profile[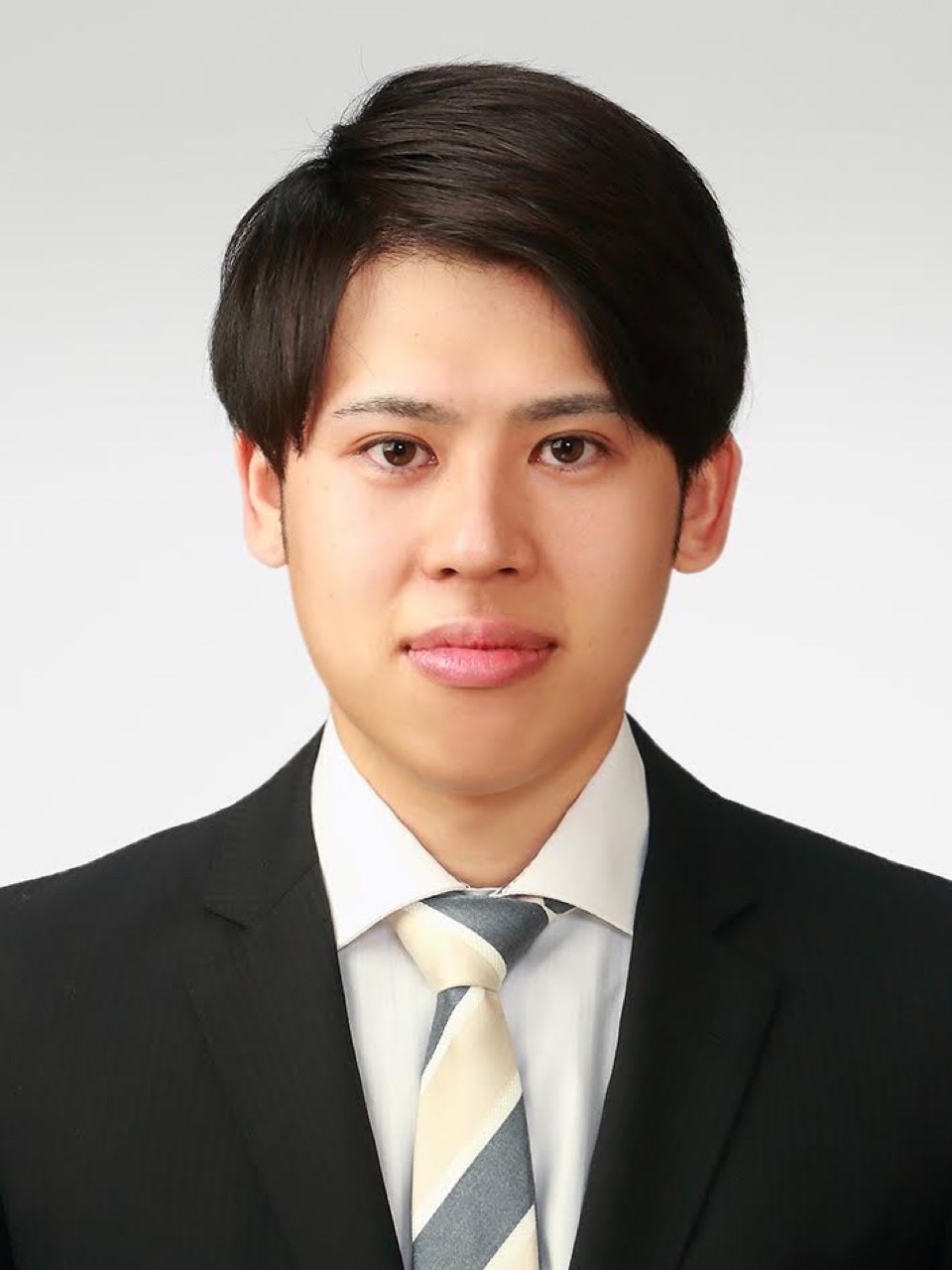]{Kota Yamashita}{received the B.E. degree in electrical and electronic engineering from Kyoto University in 2020. He is currently studying toward the M.I. degree at the Graduate School of Informatics, Kyoto University. 
    He is a member of the IEICE.}
\profile[./photo/Kamiya.jpg]{Shotaro Kamiya}{received the B.E. degree in electrical and electronic engineering from Kyoto University in 2015, and the master and Ph.D. degrees in informatics from Kyoto University in 2017 and 2020, respectively. He is currently working at Sony.}

\profile[./photo/Yamamoto.jpg]{Koji Yamamoto}{received the B.E.\ degree in electrical and electronic engineering from Kyoto University in 2002, and the master and Ph.D.\ degrees in Informatics from Kyoto University in 2004 and 2005, respectively.
From 2004 to 2005, he was a research fellow of the Japan Society for the Promotion of Science (JSPS).
Since 2005, he has been with the Graduate School of Informatics, Kyoto University, where he is currently an associate professor.
From 2008 to 2009, he was a visiting researcher at Wireless@KTH, Royal Institute of Technology (KTH) in Sweden.
He serves as an editor of IEEE Wireless Communications Letters, IEEE Open Journal of Vehicular Technology, and Journal of Communications and Information Networks, a symposium co-chair of GLOBECOM 2021, and a vice co-chair of IEEE ComSoc APB CCC.
He was a tutorial lecturer in IEEE ICC 2019.
His research interests include radio resource management, game theory, and machine learning.
He received the PIMRC 2004 Best Student Paper Award in 2004, the Ericsson Young Scientist Award in 2006.
He also received the Young Researcher's Award, the Paper Award, SUEMATSU-Yasuharu Award, Educational Service Award from the IEICE of Japan in 2008, 2011, 2016, and 2020, respectively, and IEEE Kansai Section GOLD Award in 2012.
He is a senior member of the IEEE and a member of the Operations Research Society of Japan.
}

\profile[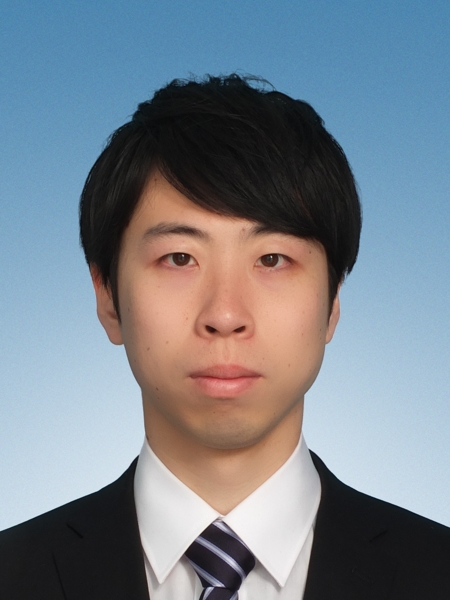]{Yusuke Koda}{(S'03--M'06) received the B.E. degree in electrical and electronic engineering from Kyoto University in 2016 and the M.E. and Ph.D. degrees (Informatics) at the Graduate School of Informatics from Kyoto University in 2018 and 2021, respectively. 
    He is currently a postdoctoral researcher at the Centre for Wireless Communications, University of Oulu. In 2019, he visited Centre for Wireless Communications, University of Oulu, Finland to conduct collaborative research. 
    He received the VTS Japan Young Researcher’s Encouragement Award in 2017 and TELECOM System Technology Award in 2020. 
    He was a Recipient of the Nokia Foundation Centennial Scholarship in 2019.
    }

\profile[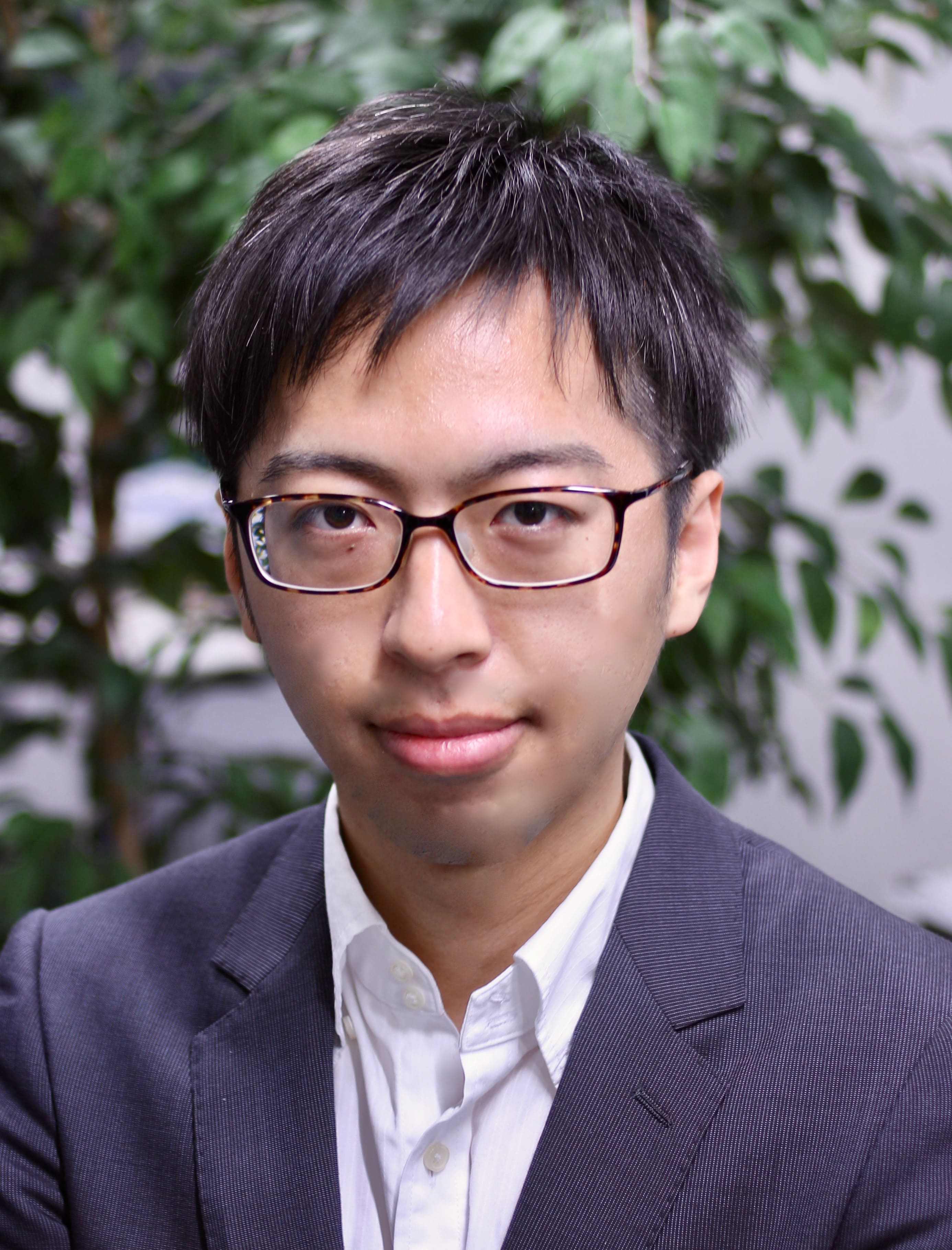]{Takayuki Nishio}{(S'11-M'14-SM'20) received the B.E.\ degree in electrical and electronic engineering and the master's and Ph.D.\ degrees in informatics from Kyoto University in 2010, 2012, and 2013, respectively. He was an assistant professor in communications and computer engineering with the Graduate School of Informatics, Kyoto University from 2013 to 2020. 
    He is currently an associate professor at the School of Engineering, Tokyo Institute of Technology, Japan.
    From 2016 to 2017, he was a visiting researcher in Wireless Information Network Laboratory (WINLAB), Rutgers University, United States. His current research interests include machine learning-based network control, machine learning in wireless networks, and heterogeneous resource management.
    }

\profile[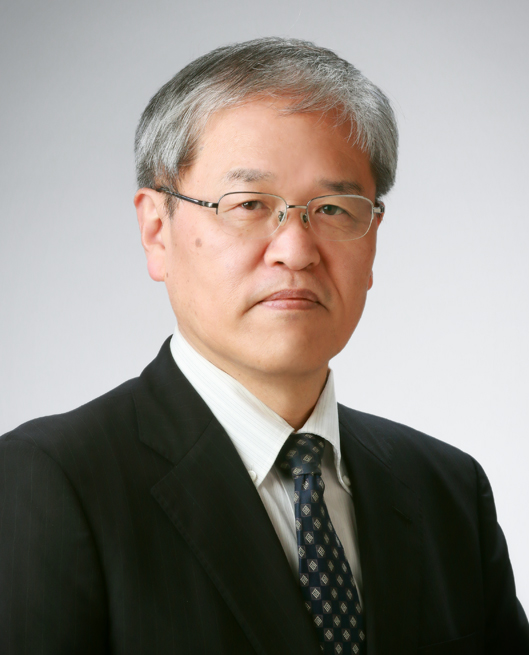]{Masahiro Morikura}{received B.E., M.E. and Ph.D. degree in electronic engineering from Kyoto University, Kyoto, Japan in 1979, 1981 and 1991, respectively. He joined NTT in 1981, where he was engaged in the research and development of TDMA equipment for satellite communications.  From 1988 to 1989, he was with the communications Research Centre, Canada as a guest scientist. From 1997 to 2002, he was active in standardization of the IEEE802.11a based wireless LAN. He received Paper Award, Achievement Award and Distinguished Achievement and Contributions Award from the IEICE in 2000, 2006 and 2019, respectively. He also received Education, Culture, Sports, Science and Technology Minister Award in 2007 and Maejima Award from the Teishin association in 2008 and the Medal of Honor with Purple Ribbon from Japan’s Cabinet Office in 2015.
      Dr. Morikura is now a professor of the Graduate School of Informatics, Kyoto University.  He is a Fellow of the IEICE and a member of IEEE.}

\end{document}